 \documentclass[final,5p,times,twocolumn,sort&compress]{elsarticle}

\usepackage [latin1]{inputenc}
\usepackage{amssymb,bm}
\usepackage{amsfonts}
\usepackage{lipsum}
\usepackage{amsmath}
\usepackage{dcolumn}
\usepackage{graphicx}
\usepackage{xcolor}
\usepackage{hyperref}
\usepackage{microtype}
\usepackage{tabularx}
\usepackage{booktabs}
\usepackage{array}

\journal{Physica E: Low-Dimensional Systems and Nanostructures}

\begin{document}

\begin{frontmatter}



\title{Global bound-state spectrum and spin- and angular-momentum-dependent confinement in an inverse-$3/2$ magnetic quantum dot}


\author[first,second]{Luis B. Castro}\ead{lrb.castro@ufma.br}
\affiliation[first]{organization={Programa de P\'{o}s-gradua\c{c}\~{a}o em F\'{i}sica, Universidade Federal do Maranh\~{a}o},
            addressline={Campus Universit\'{a}rio do Bacanga}, 
            city={S\~{a}o Lu\'{i}s},
            postcode={65080-805}, 
            state={MA},
            country={Brazil}}
\affiliation[second]{organization={Coordena\c{c}\~{a}o do Curso de F\'{i}sica - Bacharelado, Universidade Federal do Maranh\~{a}o},
            addressline={Campus Universit\'{a}rio do Bacanga}, 
            city={S\~{a}o Lu\'{i}s},
            postcode={65080-805}, 
            state={MA},
            country={Brazil}}

\begin{abstract}

We investigate the bound-state spectrum and confinement thresholds of a two-dimensional electron governed by the Schr\"{o}dinger--Pauli equation in an inverse-$3/2$ radial magnetic-field profile. The radial problem reduces to the biconfluent Heun equation, but instead of restricting the solution to the quasi-exact polynomial sector, we impose the bound-state condition globally through a spectral Wronskian between the endpoint-selected Frobenius solution and the subdominant asymptotic solution. The resulting spectrum contains continuous nonpolynomial bound-state branches on which the known quasi-exact solutions appear as isolated exact points. In this global description, the polynomial index $N$ remains a termination index within the quasi-exact sector but does not uniquely label the spectral branches. An independent exact analysis of the continuum edge yields spin- and angular-momentum-dependent critical couplings that can lie below the quasi-exact restriction $U_0a>4$, where $U_0$ is the Coulomb-strength parameter and $a$ is the magnetic-profile length scale. These results establish quasi-exact solvability as a set of special analytic points embedded in a broader biconfluent-Heun spectrum with spin- and angular-momentum-dependent confinement thresholds.
\end{abstract}



\begin{keyword}
Magnetic quantum dots \sep Spin-dependent confinement \sep Confinement thresholds \sep Bound states \sep Quasi-exact solvability \sep Biconfluent Heun equation



\end{keyword}

\end{frontmatter}




\section{Introduction}
\label{sec:introduction}

Spatially inhomogeneous magnetic fields provide a purely magnetic route to confine and manipulate carriers in two-dimensional electron systems, complementing confinement produced by electrostatic potentials. Early studies of magnetic steps, barriers, and wells showed that nonuniform perpendicular fields can support both localized and scattering states and can strongly modify the orbital dynamics of two-dimensional electrons~\cite{PRB48:15166:1993}. These ideas extend naturally to magnetic dots, antidots, and rings, where the spatial profile of the field can generate confined and magnetic-edge states with properties that differ qualitatively from those produced by conventional electrostatic confinement~\cite{PRB59:2817:1999,PR394:1:2004}. Because the orbital coupling is sensitive to angular momentum while the Zeeman interaction introduces an explicit spin dependence, such systems provide a setting in which localization can depend on both orbital and spin degrees of freedom. Recent theoretical studies have continued to explore this physics in low-dimensional quantum dots, including two-dimensional dots in perpendicular magnetic fields, magnetically driven graphene dots, and magnetic-dot spectra with explicit angular-momentum dependence~\cite{PE128:114571:2021,PE141:115245:2022,PB698:416745:2025}. Analytically tractable magnetic profiles are particularly valuable in this context because they make it possible to isolate these confinement mechanisms while retaining direct access to the underlying spectral problem.

Against this broader background, Downing and Portnoi~\cite{PRB94:045430:2016} constructed exact and quasi-exactly solvable (QES) models for two-dimensional magnetic dots and rings, including cases reducible to confluent and biconfluent Heun equations. Of particular relevance here is their inverse-$3/2$ magnetic profile, for which the radial Schr\"odinger--Pauli equation reduces to the biconfluent Heun equation (BHE) and polynomial solutions occur when two termination conditions are satisfied. Those QES solutions provide closed-form energies and wave functions at isolated parameter values and were used to discuss channel-selective vortexlike localization and spin-dependent confinement.

Polynomial truncation, however, does not generally exhaust the spectrum of the underlying eigenvalue problem. Fern\'andez~\cite{AP434:168645:2021} emphasized this point for a conditionally solvable Schr\"odinger problem governed by a three-term Frobenius recurrence: truncation selects exact polynomial states only on restricted relations among model parameters, whereas the spectral problem itself is defined more broadly. For the inverse-$3/2$ magnetic dot, this raises a corresponding set of questions: how does the bound-state spectrum evolve away from the QES parameter sets, where do its levels reach the continuum and how do those thresholds depend on spin and angular momentum, and what global spectral role, if any, is played by the polynomial index $N$?

These questions call for a global connection formulation. Global quantization strategies based on BHE connection or asymptotic coefficients have established precedents in the literature~\cite{SIAM7:157:1976,ARMA78:275:1982,AM20:87:1982,TCA133:1494:2014,NA71:797:2016}. The formulation used here brings together several established ingredients: the endpoint-selected Frobenius solution, the subdominant asymptotic solution, the Heaviside-series construction of the connection coefficients, and their reduction to a spectral Wronskian. {Wronskian-based spectral conditions have also been employed in several confluent-Heun eigenvalue problems, including rigid-rotor, Mathieu-potential, and angular-Teukolsky systems~\cite{IJQC120:e26336:2020,PLA384:126480:2020,IJQC121:e26546:2021,RIP24:104115:2021,RIP34:105330:2022,CTP74:115001:2022}. Within the present BHE formulation, the resulting bound-state condition is a scalar spectral equation that can be evaluated at generic parameter values without requiring polynomial termination.}

The purpose of this work is to determine the bound-state spectrum and confinement thresholds of the inverse-$3/2$ magnetic quantum dot without restricting the analysis to the QES sector. To this end, we formulate and compute the bound-state condition using the global BHE construction, recover the known QES states as exact benchmarks, and examine the spectrum at generic parameter values. We then treat the continuum threshold separately to characterize the onset of confinement and its dependence on spin and angular momentum. {All spectra, confinement thresholds, and parameter trends presented in this work are theoretical predictions of the model; no experimental measurements are included.}

The remainder of the paper is organized as follows. Section~\ref{sec:BHEframework} develops the global BHE framework, Sec.~\ref{sec:inverse32} applies it to the magnetic model and analyzes the bound-state spectrum and thresholds, Sec.~\ref{sec:discussion} discusses the physical and methodological implications, and Sec.~\ref{sec:conclusions} summarizes the main conclusions.

\section{Global biconfluent-Heun bound-state framework}
\label{sec:BHEframework}

We write the biconfluent Heun equation (BHE) in the canonical form~\cite{FRANK2010,NA71:797:2016}
\begin{equation}
 x y''+
 \left(1+\alpha-\beta x-2x^2\right)y'
 +\left[(\gamma-\alpha-2)x-q\right]y=0,
 \label{eq:BHE}
\end{equation}
with $q=[\delta+(1+\alpha)\beta]/2$. Equation~\eqref{eq:BHE} has a regular singular point at $x=0$ and a rank-two irregular singular point at $x=\infty$. We formulate the connection problem along the positive real axis.

The first step is to select the solution at the finite endpoint. Setting $y\sim x^p$ as $x\to0$ and retaining the terms of order $x^{p-1}$ gives the indicial equation
\begin{equation}
 p(p-1)+(1+\alpha)p=p(p+\alpha)=0.
 \label{eq:indicial}
\end{equation}
The local indices are therefore $p=0$ and $p=-\alpha$. For the bound-state construction we select the analytic branch $p=0$, denote it by $y_{\rm reg}$, and normalize it by $y_{\rm reg}(0)=1$.

\subsection{Determining the two asymptotic channels}
\label{subsec:asymptotic}

Once the solution has been fixed at $x=0$, the next question is how it can behave at the irregular endpoint. Dividing Eq.~\eqref{eq:BHE} by $x$ gives
\begin{equation}
 y''+\left[-2x-\beta+\frac{1+\alpha}{x}\right]y'
 +\left[\gamma-\alpha-2-\frac{q}{x}\right]y=0.
 \label{eq:BHEinfinity}
\end{equation}
To determine the independent large-$x$ behaviors, we use
\begin{equation}
 y(x)\sim e^{s x^2+t x}x^\sigma.
 \label{eq:asymptotic_ansatz}
\end{equation}
For this ansatz,
\begin{align}
 \frac{y'}{y}&=2sx+t+\frac{\sigma}{x},
 \label{eq:logder1}\\
 \frac{y''}{y}&=\left(2sx+t+\frac{\sigma}{x}\right)^2
 +2s-\frac{\sigma}{x^2}.
 \label{eq:logder2}
\end{align}
Substituting Eqs.~\eqref{eq:logder1} and \eqref{eq:logder2} into Eq.~\eqref{eq:BHEinfinity}, the coefficients of $x^2$ and $x$ give, respectively,
\begin{align}
 4s(s-1)&=0,
 \label{eq:sleading}\\
 (4s-2)t-2\beta s&=0.
 \label{eq:tleading}
\end{align}
The two possibilities are therefore $(s,t)=(0,0)$ and $(s,t)=(1,\beta)$. For the first pair, the constant term gives
\begin{equation}
 -2\sigma+\gamma-\alpha-2=0,
 \label{eq:sigmaeqminus}
\end{equation}
whereas for the second it gives
\begin{equation}
 2\sigma+\gamma+\alpha+2=0.
 \label{eq:sigmaeqplus}
\end{equation}
The corresponding algebraic powers are thus
\begin{equation}
 \sigma_- = \frac{\gamma-\alpha-2}{2},
 \qquad
 \sigma_+ = -\frac{\gamma+\alpha+2}{2}.
 \label{eq:sigmaBHE}
\end{equation}
The exponent $\sigma_-$ is associated with the pair $(s,t)=(0,0)$, whereas $\sigma_+$ is associated with $(s,t)=(1,\beta)$. We may therefore choose the unit-leading asymptotic solutions
\begin{align}
 Y_-(x)&\sim x^{\sigma_-}\left[1+O(x^{-1})\right],
 \label{eq:Yminus}\\
 Y_+(x)&\sim e^{x^2+\beta x}x^{\sigma_+}
 \left[1+O(x^{-1})\right].
 \label{eq:Yplus}
\end{align}
Here $Y_-$ is the algebraic BHE channel, while $Y_+$ is the independent exponential channel. The solution selected at the regular endpoint must therefore have the large-$x$ decomposition
\begin{equation}
 y_{\rm reg}(x)=A Y_-(x)+B Y_+(x).
 \label{eq:connectionBHE}
\end{equation}
The coefficients $A$ and $B$ are the regular-to-irregular connection coefficients of $y_{\rm reg}$: $A$ multiplies the algebraic channel $Y_-$, whereas $B$ multiplies the exponential channel $Y_+$.

\subsection{Converting normalizability into a spectral Wronskian}
\label{subsec:wronskian}

To formulate normalizability as a global condition, we first transform Eq.~\eqref{eq:BHE} to normal form. We introduce\cite{NA71:797:2016,EPJC79:664:2019}
\begin{equation}
 u(x)=x^{(1+\alpha)/2}
 \exp\!\left[-\frac{x^2+\beta x}{2}\right]y(x),
 \label{eq:normaltransform}
\end{equation}
which gives
\begin{equation}
 u''+\left[
 -x^2-\beta x+\gamma-\frac{\beta^2}{4}
 -\frac{\delta}{2x}+\frac{1-\alpha^2}{4x^2}
 \right]u=0.
 \label{eq:BHEnormal}
\end{equation}
For two solutions $f$ and $g$ of Eq.~\eqref{eq:BHEnormal}, we define their Wronskian by $W[f,g]\equiv fg'-f'g$. Because Eq.~\eqref{eq:BHEnormal} contains no first-derivative term, $W[f,g]$ is independent of $x$.

We denote by $u_{\rm reg}$ the image of the endpoint-selected solution $y_{\rm reg}$ under the transformation \eqref{eq:normaltransform}. Applying the same transformation to the two asymptotic channels $Y_-$ and $Y_+$ defined in Eqs.~\eqref{eq:Yminus} and \eqref{eq:Yplus}, we obtain the unit-leading Thom\'e solutions
\begin{align}
 U_-(x)&\sim e^{-(x^2+\beta x)/2}
 x^{\mu_-}\left[1+O(x^{-1})\right],
 \label{eq:Uminus}\\
 U_+(x)&\sim e^{+(x^2+\beta x)/2}
 x^{\mu_+}\left[1+O(x^{-1})\right],
 \label{eq:Uplus}
\end{align}
with $\mu_-=(\gamma-1)/2$ and $\mu_+=-(\gamma+1)/2$. Along the positive real axis, $U_-$ is the recessive channel and $U_+$ is the dominant channel. The connection formula \eqref{eq:connectionBHE} therefore becomes
\begin{equation}
 u_{\rm reg}=A U_-+B U_+.
 \label{eq:uconnection}
\end{equation}
Normalizability requires the dominant contribution to be absent, hence $B=0$.

The coefficient $B$ can now be isolated by taking the Wronskian of Eq.~\eqref{eq:uconnection} with $U_-$. Since $W[U_-,U_-]=0$,
\begin{equation}
 W[u_{\rm reg},U_-]
 =B\,W[U_+,U_-].
 \label{eq:Wconnection}
\end{equation}
The unit-leading asymptotic forms \eqref{eq:Uminus} and \eqref{eq:Uplus} give $W[U_+,U_-]\to-2$ as $x\to\infty$. Since this Wronskian is independent of $x$, it follows that
\begin{equation}
 W[U_+,U_-]=-2.
 \label{eq:Winfinity}
\end{equation}
We therefore define the spectral Wronskian
\begin{equation}
 \mathcal W\equiv W[u_{\rm reg},U_-]=-2B.
 \label{eq:spectralW}
\end{equation}
Consequently, the normalizability condition $B=0$ is equivalent to
\begin{equation}
 \mathcal W=0.
 \label{eq:normalizability}
\end{equation}
Thus the bound-state condition is expressed as the linear dependence of the solution selected at the regular endpoint and the recessive Thom\'e solution selected at infinity.

\subsection{Computing the Wronskian from recurrence data}
\label{subsec:recurrence}

Equation~\eqref{eq:normalizability} reduces the bound-state problem to the condition $\mathcal W=0$, but the two solutions entering $\mathcal W=W[u_{\rm reg},U_-]$ are specified at opposite endpoints. We therefore need a way to evaluate the constant Wronskian from coefficient data associated with those two endpoint selections, without propagating either solution across the full interval. We use the BHE connection construction of Ferreira and Sesma~\cite{NA71:797:2016}, which implements the Heaviside-series method of Naundorf~\cite{SIAM7:157:1976}.

We first construct the coefficient sequence of the recessive Thom\'e solution. We set $\mu\equiv\mu_-=(\gamma-1)/2$ and write
\begin{equation}
 U_-(x)\sim e^{-(x^2+\beta x)/2}
 x^\mu\sum_{n=0}^{\infty}a_n x^{-n}.
 \label{eq:ThomeShort}
\end{equation}
We choose the unit-leading normalization $a_0=1$ and set $a_{-1}=0$. Substituting Eq.~\eqref{eq:ThomeShort} into the normal-form equation \eqref{eq:BHEnormal} then gives, for $n\geq1$,
\begin{align}
 2n a_n={}&
 \left[\frac{\delta}{2}-\beta(n-1-\mu)\right]a_{n-1}
 \nonumber\\
 &-\left[(n-2-\mu)(n-1-\mu)
 +\frac{1-\alpha^2}{4}\right]a_{n-2}.
 \label{eq:Thomerec}
\end{align}
Thus Eq.~\eqref{eq:Thomerec} determines the asymptotic coefficients $a_n$ of $U_-$.

We next construct the coefficient sequence associated with the solution selected at $x=0$. We set $\rho\equiv(1+\alpha)/2$ and define $\widehat u(x)\equiv e^{-\beta x/2}u_{\rm reg}(x)$. Near the regular endpoint this function has the Frobenius expansion
\begin{equation}
 \widehat u(x)=x^\rho\sum_{n=0}^{\infty}\widehat c_n x^n,
 \label{eq:uhat}
\end{equation}
where $\widehat c_0=1$ and $\widehat c_n=0$ for $n<0$. Since $u_{\rm reg}=e^{\beta x/2}\widehat u$, substitution into Eq.~\eqref{eq:BHEnormal} gives
\begin{equation}
 \widehat u''+\beta\widehat u'
 +\left[
 -x^2-\beta x+\gamma
 -\frac{\delta}{2x}+\frac{1-\alpha^2}{4x^2}
 \right]\widehat u=0.
 \label{eq:uhateq}
\end{equation}
Substituting the series \eqref{eq:uhat} into Eq.~\eqref{eq:uhateq} and equating equal powers of $x$ yields
\begin{align}
 n(n+2\rho-1)\widehat c_n={}&
 \left[\frac{\delta}{2}-\beta(n+\rho-1)\right]
 \widehat c_{n-1}
 -\gamma\widehat c_{n-2}
 \nonumber\\
 &+\beta\widehat c_{n-3}+\widehat c_{n-4}.
 \label{eq:chatrec}
\end{align}
Equation~\eqref{eq:chatrec} therefore determines the coefficients $\widehat c_n$ of the endpoint-selected solution in the form needed below.

Having obtained both coefficient sequences, we now rewrite the spectral Wronskian $\mathcal W=W[u_{\rm reg},U_-]$ defined in Eq.~\eqref{eq:spectralW} in a form suitable for coefficient matching. We multiply both solutions entering $\mathcal W$ by the common factor $e^{x^2/4}$ and define
\begin{equation}
 v(x)\equiv e^{x^2/4}u_{\rm reg}(x),
 \qquad
 v_-(x)\equiv e^{x^2/4}U_-(x).
 \label{eq:vdefs}
\end{equation}
For any common factor $f(x)$, the Wronskian satisfies $W[fu,fg]=f^2W[u,g]$. Therefore,
\begin{equation}
 W[v,v_-]=e^{x^2/2}\mathcal W.
 \label{eq:Waux}
\end{equation}
If we denote the algebraic series in Eq.~\eqref{eq:ThomeShort} by $S_-(x)\equiv x^\mu\sum_{m=0}^{\infty}a_m x^{-m}$, then Eqs.~\eqref{eq:ThomeShort}, \eqref{eq:uhat}, and \eqref{eq:vdefs} give
\begin{equation}
 v=e^{x^2/4+\beta x/2}\widehat u,
 \qquad
 v_-=e^{-x^2/4-\beta x/2}S_-.
 \label{eq:vforms}
\end{equation}
The exponential factors cancel when the Wronskian is formed, and Eq.~\eqref{eq:Waux} has the left-hand side
\begin{equation}
 W[v,v_-]=-(x+\beta)\widehat u S_-
 +\widehat u S_-'-\widehat u' S_-.
 \label{eq:Wexpanded}
\end{equation}
Substituting Eq.~\eqref{eq:uhat} for $\widehat u$ and $S_-(x)=x^\mu\sum_{m=0}^{\infty}a_m x^{-m}$ into Eq.~\eqref{eq:Wexpanded} yields the bilateral expansion
\begin{equation}
 W[v,v_-]\sim
 \sum_{s=-\infty}^{\infty}g_s x^{s+\rho+\mu},
 \label{eq:gsseries}
\end{equation}
where the coefficient of each power is
\begin{align}
 g_s=\sum_{m=0}^{\infty}a_m\Big[&
 -\widehat c_{s+m-1}
 -\beta\widehat c_{s+m}
 \nonumber\\
 &+(\mu-\rho-s-2m-1)\widehat c_{s+m+1}\Big].
 \label{eq:gs}
\end{align}

It remains to match this bilateral power series to the factor $e^{x^2/2}$ on the right-hand side of Eq.~\eqref{eq:Waux}. For that purpose we use Heaviside's expansion~\cite{HARDY1967,SIAM7:157:1976}, valid for $|\arg t|<\pi$,
\begin{equation}
 e^t\sim\sum_{n=-\infty}^{\infty}
 \frac{t^{n+\Delta}}{\Gamma(n+1+\Delta)}.
 \label{eq:Heaviside}
\end{equation}
With $t=x^2/2$, the powers in Eq.~\eqref{eq:gsseries} separate into the two parity classes $s=2\mathcal N+L$, where $L=0$ or $1$. For each class we define $\Delta_L\equiv(\rho+\mu+L)/2$. Matching the corresponding powers in Eqs.~\eqref{eq:gsseries} and \eqref{eq:Heaviside} to Eq.~\eqref{eq:Waux} gives
\begin{equation}
 \mathcal W=
 \sum_{L=0}^{1}
 \frac{\Gamma(\mathcal N+1+\Delta_L)}
 {(1/2)^{\mathcal N+\Delta_L}}
 g_{2\mathcal N+L}.
 \label{eq:Wrecurrence}
\end{equation}
Here $\mathcal N$ is an auxiliary integer chosen so that the representation is regular. The exact value of $\mathcal W$ is independent of that choice; numerically, varying $\mathcal N$ therefore provides an internal stability test, together with convergence of the sums defining $g_s$. Combining Eq.~\eqref{eq:Wrecurrence} with the bound-state condition \eqref{eq:normalizability} turns the global connection problem into a root problem constructed entirely from recurrence data.

As a check on conventions, for $\alpha=-1$ and $\gamma=1$, Eqs.~\eqref{eq:Thomerec}, \eqref{eq:chatrec}, and \eqref{eq:Wrecurrence} reduce to the BHE Wronskian construction used in Ref.~\cite{EPJC79:664:2019}.

\subsection{Biconfluent-Heun polynomial sector of the global connection problem}
\label{sec:polynomial}

The global condition $\mathcal W=0$ does not require the solution selected at $x=0$ to be a polynomial. We now identify the polynomial solutions that occur as a special sector of the same boundary-value problem.

The analytic solution $y_{\rm reg}$ selected at the regular endpoint has the Frobenius expansion
\begin{equation}
 y_{\rm reg}(x)=\sum_{n=0}^{\infty}d_n x^n,
 \label{eq:frobenius}
\end{equation}
with $d_{-1}=0$ and $d_0=1$. Substituting this series into the canonical BHE \eqref{eq:BHE} and equating equal powers of $x$ gives, for $n\geq0$,
\begin{equation}
 (n+1)(n+1+\alpha)d_{n+1}
 = (\beta n+q)d_n+(2n+\alpha-\gamma)d_{n-1}.
 \label{eq:BHErec}
\end{equation}

Suppose the series terminates at degree $N$, so that $d_N\neq0$ and $d_n=0$ for every $n>N$. The recurrence at $n=N+1$ can then be satisfied only if
\begin{equation}
 \gamma-\alpha-2=2N,
 \qquad N=0,1,2,\ldots,
 \label{eq:BHEpoly1}
\end{equation}
while termination itself requires
\begin{equation}
 d_{N+1}=0.
 \label{eq:BHEpoly2}
\end{equation}
These are the two independent polynomial conditions for the BHE~\cite{TCA133:1494:2014,PRB94:045430:2016}. Once both hold, the coefficient multiplying $d_{n-1}$ in Eq.~\eqref{eq:BHErec} vanishes at $n=N+1$, and the recurrence forces all coefficients beyond $d_N$ to vanish.

The first polynomial condition also identifies the asymptotic channel of the terminated solution. Substituting Eq.~\eqref{eq:BHEpoly1} into the exponent $\sigma_-=(\gamma-\alpha-2)/2$ derived in Eq.~\eqref{eq:sigmaBHE} gives $\sigma_-=N$. Thus a degree-$N$ polynomial has the same leading algebraic power as the canonical channel $Y_-$ in Eq.~\eqref{eq:Yminus}. Because a polynomial cannot contain the exponentially growing channel $Y_+$ of Eq.~\eqref{eq:Yplus} along the positive real axis, a polynomial that is also the endpoint-selected solution $y_{\rm reg}$ has $B=0$ and therefore satisfies $\mathcal W=0$.

The converse does not follow. The global normalizability condition $\mathcal W=0$ removes the dominant asymptotic channel, but it does not require the Frobenius series \eqref{eq:frobenius} to terminate. The polynomial solutions are therefore analytic benchmark points inside the unrestricted global bound-state spectrum, not the general quantization condition.

\section{Inverse-3/2 magnetic quantum dot}
\label{sec:inverse32}

We now apply the global construction to the inverse-$3/2$ magnetic model introduced by Downing and Portnoi~\cite{PRB94:045430:2016}. The purpose is first to identify the endpoint-selected BHE associated with this problem and its global bound-state condition, then to compare the known quasi-exact polynomial states with the unrestricted spectrum selected by the spectral Wronskian, and finally to examine the continuum threshold and its dependence on spin and angular momentum.

\subsection{Model and global bound-state condition}
\label{subsec:inverse32_model}

Consider an electron of mass $M$ and charge $-e$, with $e>0$, confined to the plane. We denote its canonical momentum by $\bm p$, the vector potential by $\bm A$, the scalar potential by $V(r)$, and the perpendicular magnetic field by $B_z(r)$. Writing $\mu_B=e\hbar/(2M)$ for the Bohr magneton, a stationary state $\Psi_\tau$ of energy $E$ and spin label $\tau=\pm1$ obeys the Schr\"odinger--Pauli equation
\begin{equation}
 \left[
 \frac{(\bm p+e\bm A)^2}{2M}
 +\tau\mu_B B_z(r)+V(r)
 \right]\Psi_\tau=E\Psi_\tau.
 \label{eq:Pauli_inverse32}
\end{equation}
Following Ref.~\cite{PRB94:045430:2016}, we introduce the rescaled energy $\epsilon=2ME/\hbar^2$ and potential $U(r)=2MV(r)/\hbar^2$, and take the attractive Coulomb potential $U(r)=-U_0/r$ with $U_0>0$. The inverse-$3/2$ magnetic profile is
\begin{equation}
 B_z(r)=\frac{\hbar}{e}\frac{1}{a^{1/2}r^{3/2}},
 \label{eq:Bfield_inverse32}
\end{equation}
where $a$ is a length scale. In the azimuthal gauge $\bm A=A_\theta(r)\,\hat{\bm\theta}$, the field satisfies $B_z(r)=r^{-1}d[rA_\theta(r)]/dr$. Equation~\eqref{eq:Bfield_inverse32} is therefore generated for $r>0$ by the choice $A_\theta(r)=2\hbar/[e\sqrt{ar}]$, which we adopt below.

Because the problem is rotationally symmetric, we write $\Psi_\tau(r,\theta)=(2\pi)^{-1/2}e^{im\theta}\psi_\tau(r)$, where the integer $m$ is the azimuthal quantum number. Substituting this form and the potentials above into Eq.~\eqref{eq:Pauli_inverse32}, and then introducing the dimensionless radius $\xi=r/a$ and energy $\mathcal E=\epsilon a^2$, gives
\begin{equation}
 \psi_\tau''+\frac{1}{\xi}\psi_\tau'
 +\left[
 \mathcal E+\frac{U_0a-4}{\xi}
 -\frac{4m+\tau}{\xi^{3/2}}
 -\frac{m^2}{\xi^2}
 \right]\psi_\tau=0,
 \label{eq:radial_inverse32}
\end{equation}
where primes denote derivatives with respect to $\xi$.

The allowed energy range follows from the large-$\xi$ limit of Eq.~\eqref{eq:radial_inverse32}. For $\mathcal E\neq0$, the inverse powers of $\xi$ vanish asymptotically, so the leading equation is $\psi_\tau''+\xi^{-1}\psi_\tau'+\mathcal E\psi_\tau\simeq0$. For $\mathcal E>0$ the asymptotic solutions are oscillatory, whereas $\mathcal E<0$ gives exponentially decreasing and increasing branches. Thus $\mathcal E=0$ is the continuum threshold and must be treated separately, since the leading asymptotic balance changes there. In the ordinary bound-state sector $\mathcal E<0$, we parameterize the energy as $\mathcal E=-\kappa^2$, with $\kappa>0$.

The behavior at the origin is fixed independently. Setting $\psi_\tau\sim\xi^p$ in Eq.~\eqref{eq:radial_inverse32}, the terms of order $\xi^{p-2}$ give $p^2-m^2=0$. With $\ell\equiv|m|$, the two leading powers are $\xi^{\ell}$ and $\xi^{-\ell}$ for $\ell>0$, and regularity selects $\xi^{\ell}$; for $m=0$ the exponent $p=0$ is repeated. At infinity, normalizability selects the branch with exponential factor $e^{-\kappa\xi}$. To remove the selected origin power and asymptotic exponential before identifying the remaining equation, we introduce $x=(2\kappa\xi)^{1/2}$ and write
\begin{equation}
 \psi_\tau(\xi)=x^{2\ell}e^{-x^2/2}y(x),
 \label{eq:factor_inverse32}
\end{equation}
up to an overall normalization. Here $x^{2\ell}\propto\xi^\ell$ and $e^{-x^2/2}=e^{-\kappa\xi}$.

Substituting Eq.~\eqref{eq:factor_inverse32}, together with $\xi=x^2/(2\kappa)$, into Eq.~\eqref{eq:radial_inverse32} gives the canonical BHE~\eqref{eq:BHE} with
\begin{align}
 \alpha&=4\ell,
 &\beta&=0,
 \label{eq:ab_inverse32}\\
 \gamma&=\frac{2(U_0a-4)}{\kappa},
 &\delta&=\frac{4\sqrt{2}(4m+\tau)}{\sqrt{\kappa}}.
 \label{eq:gd_inverse32}
\end{align}
This is the BHE parametrization of the inverse-$3/2$ model written in the convention used throughout Sec.~\ref{sec:BHEframework}~\cite{PRB94:045430:2016}.

We can now identify the physical endpoint solution within the generic construction of Sec.~\ref{sec:BHEframework}. The local Frobenius exponents at $x=0$ obtained from Eq.~\eqref{eq:indicial} are $p=0$ and $p=-\alpha$. With Eq.~\eqref{eq:ab_inverse32}, they become $p=0$ and $p=-4\ell$. Restoring the prefactor $x^{2\ell}$ in Eq.~\eqref{eq:factor_inverse32}, these two BHE branches give $\psi_\tau\sim r^\ell$ and $\psi_\tau\sim r^{-\ell}$ for $\ell>0$. Regularity therefore selects the analytic branch $p=0$, namely the endpoint-selected solution $y_{\rm reg}$ of Sec.~\ref{sec:BHEframework}. When $m=0$, the two exponents coincide and regularity again selects the analytic solution rather than its logarithmic companion. {In this channel, $\alpha=0$ and $\rho=1/2$, so the factor $n(n+2\rho-1)$ in Eq.~\eqref{eq:chatrec} reduces to $n^2$, which is nonzero for every $n\geq1$. The recurrence construction therefore remains well defined, and no restriction to $m\neq0$ is required.}

The asymptotic selection follows from the same factorization. Since $\beta=0$, Eqs.~\eqref{eq:Yminus} and \eqref{eq:Yplus} give $Y_-\sim x^{\sigma_-}$ and $Y_+\sim e^{x^2}x^{\sigma_+}$. Multiplication by the common factor $x^{2\ell}e^{-x^2/2}$ in Eq.~\eqref{eq:factor_inverse32} therefore makes the $Y_-$ contribution decay as $e^{-\kappa\xi}$, up to an algebraic power, while the $Y_+$ contribution grows as $e^{+\kappa\xi}$. Thus $Y_-$ is the normalizable channel and $Y_+$ is the inadmissible growing channel.

For fixed $m$, $\tau$, and $U_0a$, Eqs.~\eqref{eq:ab_inverse32} and \eqref{eq:gd_inverse32} make the spectral Wronskian defined in Eq.~\eqref{eq:spectralW} a function of $\kappa$. We denote this specialized function by $\mathcal W_{m\tau}(\kappa;U_0a)$ and evaluate it from the recurrence representation~\eqref{eq:Wrecurrence}. The bound-state condition is therefore
\begin{equation}
 \mathcal W_{m\tau}(\kappa;U_0a)=0.
 \label{eq:global_inverse32}
\end{equation}
Each positive root $\kappa$ determines the dimensionless energy through $\mathcal E=-\kappa^2$.

\subsection{QES benchmarks and nonpolynomial bound states}
\label{subsec:inverse32_spectrum}

The polynomial sector identified in Sec.~\ref{sec:polynomial} provides a direct benchmark for the global condition~\eqref{eq:global_inverse32}. For a degree-$N$ polynomial, substituting Eqs.~\eqref{eq:ab_inverse32} and \eqref{eq:gd_inverse32} into the first termination condition~\eqref{eq:BHEpoly1} gives
$\kappa=(U_0a-4)/(1+N+2|m|)$. Since $\kappa>0$ in the bound-state sector and the denominator is positive, a polynomial bound state requires $U_0a>4$. Using $\mathcal E=-\kappa^2$, the corresponding candidate energy is
\begin{equation}
 \mathcal E^{\rm QES}_{N,m}
 =-\left[
 \frac{U_0a-4}{1+N+2|m|}
 \right]^2.
 \label{eq:QES_inverse32}
\end{equation}
This reproduces the first QES condition of Ref.~\cite{PRB94:045430:2016}. It is not sufficient by itself: a degree-$N$ BHE polynomial must also satisfy the second termination condition $d_{N+1}=0$.

The case $N=0$ illustrates the role of the second condition explicitly. Setting $n=0$ in the Frobenius recurrence~\eqref{eq:BHErec}, with $d_{-1}=0$ and $d_0=1$, gives $(1+\alpha)d_1=q$. For the present model $\beta=0$, so $q=\delta/2$ and hence $d_1=\delta/[2(1+\alpha)]$. A degree-zero polynomial would therefore require $\delta=0$. For $m=1$, Eq.~\eqref{eq:gd_inverse32} gives $4m+\tau=5$ for $\tau=+1$ and $4m+\tau=3$ for $\tau=-1$, so $\delta$ cannot vanish. The first polynomial state in either spin family consequently occurs at $N=1$.

\begin{figure*}[t]
 \includegraphics[width=0.45\textwidth]{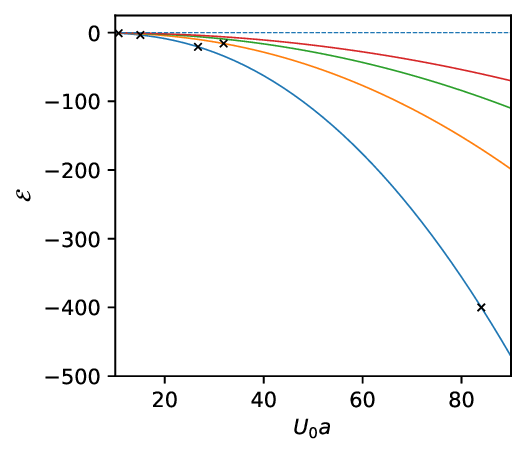}\hfill
 \includegraphics[width=0.45\textwidth]{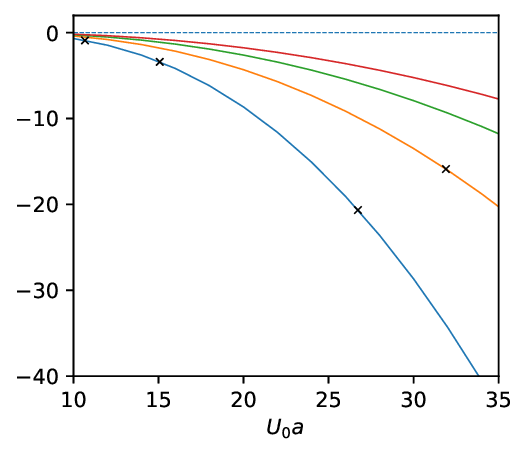}
 \caption{Dimensionless energy $\mathcal E=\epsilon a^2$ versus $U_0a$ for $m=1$ and $\tau=+1$. The left panel shows the four lowest-energy roots of $\mathcal W_{1,+1}=0$ over the full displayed interval, and the right panel enlarges the region $10\leq U_0a\leq35$. Solid curves denote the global bound-state branches, black crosses mark BHE-polynomial points satisfying both termination conditions for $N=1,2,3,4$, and the dashed horizontal line denotes the continuum threshold $\mathcal E=0$.}
 \label{fig:inverse32_branches}
\end{figure*}

We now use the $m=1$ polynomial states as benchmarks of Eq.~\eqref{eq:global_inverse32}. Solving $d_{N+1}=0$ together with the relation for $\kappa$ obtained above gives, for $N=1$, $(U_0a,\mathcal E)=(84,-400)$ when $\tau=+1$ and $(32.8,-51.84)$ when $\tau=-1$. For $N=2$, the corresponding pairs are $(26.7272727,-20.6611570)$ and $(12.1818182,-2.67768595)$, respectively. These values reproduce the results reported by Downing and Portnoi~\cite{PRB94:045430:2016}. Evaluating the specialized Wronskian through Eq.~\eqref{eq:Wrecurrence} at the same parameter pairs gives $\mathcal W_{m\tau}=0$ to numerical accuracy, as required by the general result of Sec.~\ref{sec:polynomial}.

To exhibit a bound state outside the polynomial sector, we keep $m=1$ and $\tau=+1$ fixed and set $U_0a=30$. Solving the global condition~\eqref{eq:global_inverse32}, with the Wronskian evaluated from Eq.~\eqref{eq:Wrecurrence}, gives the lowest-energy root
\begin{equation}
 \kappa=5.3551924733,
 \qquad
 \mathcal E=-28.6780864258.
 \label{eq:nonpoly_inverse32}
\end{equation}
For this root, the recessive-channel exponent defined in Eq.~\eqref{eq:sigmaBHE} is $\sigma_-=1.8551009380$. Since polynomial termination requires $\sigma_-=N\in\mathbb N_0$, this state is nonpolynomial. The numerical value of $\kappa$ is unchanged at the displayed precision when the auxiliary index in Eq.~\eqref{eq:Wrecurrence} is varied over $\mathcal N=8,10,12,14$ and the recurrence truncation over $100,140,180,220$. {The Wronskian root search was performed using high-precision arithmetic with an internal convergence tolerance of $10^{-35}$. As an independent validation in the nonpolynomial regime, direct numerical integration of the radial equation was carried out for representative values $U_0a=10,20,30,$ and $35$, with the corresponding roots determined independently to a tolerance of $10^{-14}$. The absolute differences between the Wronskian and direct-integration values of $\kappa$ are $4.7\times10^{-15}$, $1.5\times10^{-13}$, $3.2\times10^{-14}$, and $8.0\times10^{-14}$, respectively. Thus, over the representative nonpolynomial points considered, the two methods agree within $1.6\times10^{-13}$ in $\kappa$.}

Figure~\ref{fig:inverse32_branches} places the global spectrum in a broader parameter range. At fixed $m=1$ and $\tau=+1$, the four lowest-energy zeros of $\mathcal W_{1,+1}$ form continuous branches as $U_0a$ varies. The BHE-polynomial solutions occur only at isolated points where both termination conditions are satisfied, while the neighboring global roots persist without termination of the Frobenius series. In particular, the two degree-$N=4$ polynomial points visible in the plotted interval lie on different global branches. The polynomial degree is therefore a termination index, not a label of a unique bound-state branch.

\subsection{Exact confinement thresholds and spin--angular-momentum selection}
\label{subsec:inverse32_thresholds}

The polynomial benchmarks of Sec.~\ref{subsec:inverse32_spectrum} all satisfy $U_0a>4$, as required by the first termination condition, but this restriction belongs to the QES sector rather than to the global bound-state condition. To determine where confinement actually begins, we return to the continuum edge $\mathcal E=0$. In the subcritical regime $U_0a<4$, define $\zeta=\sqrt{\xi}$ and $\lambda=\sqrt{4-U_0a}>0$. Equation~\eqref{eq:radial_inverse32} then becomes
\begin{equation}
 \frac{d^2\psi_\tau}{d\zeta^2}+\frac{1}{\zeta}\frac{d\psi_\tau}{d\zeta}
 +\left[-4\lambda^2-\frac{4(4m+\tau)}{\zeta}-\frac{4m^2}{\zeta^2}\right]\psi_\tau=0.
 \label{eq:threshold_s}
\end{equation}
Introducing $z=4\lambda\zeta$ and factoring
$\psi_\tau=z^{2|m|}e^{-z/2}f(z)$ gives
\begin{equation}
 z f''+(4|m|+1-z)f'
 -\left(2|m|+\frac{1}{2}+\frac{4m+\tau}{\lambda}\right)f=0,
 \label{eq:threshold_kummer}
\end{equation}
which is Kummer's equation~\cite{FRANK2010}. With
$a_{\rm K}=2|m|+\tfrac{1}{2}+(4m+\tau)/\lambda$ and $b_{\rm K}=4|m|+1$, the solution regular at the origin is
$f(z)=M(a_{\rm K},b_{\rm K};z)\equiv{}_1F_1(a_{\rm K};b_{\rm K};z)$. {For $z\to+\infty$, and for $a_{\rm K}$ not a nonpositive integer, the regular Kummer solution has the asymptotic form
\begin{equation}
M(a_{\rm K},b_{\rm K};z)\sim
\frac{\Gamma(b_{\rm K})}{\Gamma(a_{\rm K})}
e^z z^{a_{\rm K}-b_{\rm K}}
+
\frac{\Gamma(b_{\rm K})}{\Gamma(b_{\rm K}-a_{\rm K})}
e^{\pm i\pi a_{\rm K}}z^{-a_{\rm K}}.
\label{eq:kummer_asymptotic_revision}
\end{equation}
After multiplication by the prefactor $e^{-z/2}$, the first term produces an inadmissible contribution proportional to $e^{z/2}z^{a_{\rm K}-b_{\rm K}}$, whereas the second gives a contribution proportional to $e^{-z/2}z^{-a_{\rm K}}$. The growing contribution is absent precisely when $1/\Gamma(a_{\rm K})=0$, that is,
\begin{equation}
a_{\rm K}=-n_r,\qquad n_r=0,1,2,\ldots .
\label{eq:kummer_termination_revision}
\end{equation}
For these values the Kummer series terminates exactly,
\begin{equation}
M(-n_r,b_{\rm K};z)
=
\frac{n_r!}{(b_{\rm K})_{n_r}}
L_{n_r}^{(b_{\rm K}-1)}(z),
\label{eq:kummer_laguerre_revision}
\end{equation}
so that the threshold solution becomes an exponentially decaying factor multiplied by a polynomial. Substituting $a_{\rm K}=-n_r$ into its definition then gives}
\begin{equation}
 2|m|+\frac{1}{2}+\frac{4m+\tau}{\sqrt{4-U_0a}}=-n_r.
 \label{eq:threshold_quantization}
\end{equation}
A real solution of Eq.~\eqref{eq:threshold_quantization} requires $4m+\tau<0$. The corresponding critical couplings are therefore
\begin{equation}
 (U_0a)_{c,n_r}^{(m,\tau)}
 =4-\frac{(4m+\tau)^2}
 {\left(n_r+2|m|+\tfrac{1}{2}\right)^2}.
 \label{eq:critical_couplings}
\end{equation}
{Using the standard relation $M(-n,b;z)=n!L_n^{(b-1)}(z)/(b)_n$~\cite{FRANK2010}, the threshold state at these couplings can be written, up to a normalization constant $\mathcal C$, as
\begin{equation}
\psi_\tau(z)=\mathcal C\,z^{2|m|}e^{-z/2}L_{n_r}^{(4|m|)}(z).
\label{eq:threshold_state_revision}
\end{equation}
Its square integrability can be verified directly with the radial measure. Since $z=4\lambda\sqrt{\xi}$ and $\xi=r/a$,
\begin{equation}
r\,dr=\frac{a^2}{128\lambda^4}\,z^3\,dz,
\label{eq:radial_measure_revision}
\end{equation}
and therefore
\begin{equation}
\int_0^\infty r\,|\psi_\tau(r)|^2\,dr
=
\frac{a^2|\mathcal C|^2}{128\lambda^4}
\int_0^\infty
z^{4|m|+3}e^{-z}
\left[L_{n_r}^{(4|m|)}(z)\right]^2dz.
\label{eq:threshold_norm_revision}
\end{equation}
Near the origin the integrand behaves as $z^{4|m|+3}$, while at large $z$ it is a polynomial times $e^{-z}$. The integral is therefore finite for every $m\in\mathbb Z$ and $n_r\in\mathbb N_0$. In particular, for $n_r=0$ and $m=0$, $L_0^{(0)}(z)=1$ and the remaining integral is $\int_0^\infty z^3e^{-z}\,dz=\Gamma(4)=6$, which is finite. Consequently, the continuum edge itself contains normalizable states at the isolated parameter values given by Eq.~\eqref{eq:critical_couplings}.}

Equation~\eqref{eq:critical_couplings} gives a direct spin--angular-momentum selection rule. For $\tau=+1$, subcritical threshold states require $m\leq-1$, whereas for $\tau=-1$ they occur for $m\leq0$. For $\tau=-1$ and $m\leq0$, the lowest threshold is
\begin{equation}
 (U_0a)_{c,0}^{(m,-1)}=0,
 \label{eq:zero_threshold_minus}
\end{equation}
while for $m<0$ and $\tau=+1$,
\begin{equation}
 (U_0a)_{c,0}^{(m,+1)}
 =\frac{64|m|}{(4|m|+1)^2}>0.
 \label{eq:zero_threshold_plus}
\end{equation}
{Thus, in the subcritical regime $U_0a<4$, bound states can occur only in channels satisfying $4m+\tau<0$, and only once $U_0a$ exceeds the corresponding lowest threshold $(U_0a)_{c,0}^{(m,\tau)}$. Channels with $4m+\tau\geq0$ do not support subcritical bound states.}
The ordering of the thresholds also determines the number of negative-energy states below $U_0a=4$. To establish the monotonicity required for this counting, set $\chi(\xi)=\sqrt{\xi}\,\psi_\tau(\xi)$. Equation~\eqref{eq:radial_inverse32} can then be written as a one-dimensional eigenvalue problem with effective Hamiltonian
\begin{equation}
 H_{m\tau}=-\frac{d^2}{d\xi^2}
 +\frac{m^2-\tfrac14}{\xi^2}
 +\frac{4m+\tau}{\xi^{3/2}}
 +\frac{4-U_0a}{\xi}.
 \label{eq:effective_hamiltonian}
\end{equation}
For a normalized discrete state, the Hellmann--Feynman relation~\cite{HELLMANN1937,PR56:340:1939} gives
$d\mathcal E/d(U_0a)=-\langle\xi^{-1}\rangle<0$. Each level therefore crosses the continuum edge only once as $U_0a$ increases. {For a fixed $U_0a<4$, the level labeled by $n_r$ has entered the negative-energy sector when $U_0a>(U_0a)_{c,n_r}^{(m,\tau)}$. Using Eq.~\eqref{eq:critical_couplings}, this condition is equivalent to
\begin{equation}
 n_r<
 \frac{|4m+\tau|}{\sqrt{4-U_0a}}
 -2|m|-\frac{1}{2}.
 \label{eq:bound_state_count_condition_revision}
\end{equation}
The inequality is strict because a state exactly at its critical coupling has $\mathcal E=0$ and is not included in the negative-energy count. Since $n_r=0,1,2,\ldots$, the number of nonnegative integers satisfying this inequality is therefore}
\begin{equation}
 N_{\rm b}^{(m,\tau)}
 =\max\!\left\{0,
 \left\lceil
 \frac{|4m+\tau|}{\sqrt{4-U_0a}}
 -2|m|-\frac{1}{2}
 \right\rceil\right\}.
 \label{eq:bound_state_count}
\end{equation}
For example, at $m=-1$ the first three thresholds for $\tau=+1$ are $2.56$, $3.265306\ldots$, and $3.555556\ldots$, whereas for $\tau=-1$ they are $0$, $1.959184\ldots$, and $2.765432\ldots$. At $U_0a=3$, Eq.~\eqref{eq:bound_state_count} consequently gives one negative-energy state for $\tau=+1$ and three for $\tau=-1$.

\begin{figure*}[t]
 \includegraphics[width=0.45\textwidth]{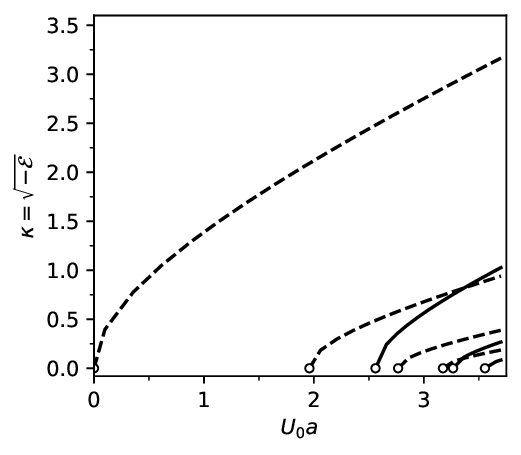}\hfill
 \includegraphics[width=0.45\textwidth]{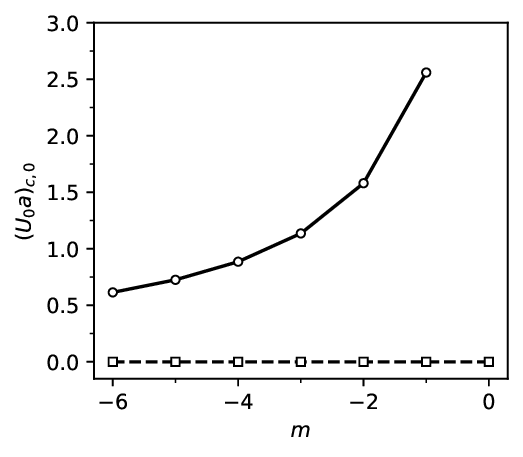}
 \caption{Subcritical confinement for the inverse-$3/2$ model. Left panel: positive Wronskian roots $\kappa=\sqrt{-\mathcal E}$ for $m=-1$ and $U_0a<4$. Solid curves correspond to $\tau=+1$ and dashed curves to $\tau=-1$; open circles mark the exact zero-energy thresholds from Eq.~\eqref{eq:critical_couplings}. The first three $\tau=+1$ branches and first four $\tau=-1$ branches are shown. Right panel: lowest critical coupling $(U_0a)_{c,0}$ versus $m$. Solid circles denote $\tau=+1$ and dashed squares denote $\tau=-1$ for the channels satisfying $4m+\tau<0$.}
 \label{fig:inverse32_thresholds}
\end{figure*}

Figure~\ref{fig:inverse32_thresholds} connects the exact threshold analysis to the global BHE condition. The left panel follows zeros of $\mathcal W_{-1,\tau}(\kappa;U_0a)$ from the negative-energy sector to the critical couplings where $\kappa\to0$. The two spin orientations have visibly different onset sequences, while every displayed branch terminates at the corresponding value of Eq.~\eqref{eq:critical_couplings}. The right panel shows that this distinction persists across angular-momentum channels: the $\tau=-1$ lowest thresholds remain at $U_0a=0$ for $m\leq0$, whereas the $\tau=+1$ thresholds are finite for $m<0$ and decrease with increasing $|m|$.

\section{Discussion}
\label{sec:discussion}

The global bound-state spectrum of the inverse-$3/2$ magnetic quantum dot places quasi-exact solvability in a broader confinement picture. In the original QES construction, polynomiality singles out isolated parameter sets for which the local Frobenius series terminates~\cite{PRB94:045430:2016}. Within the present formulation, those states remain exact analytic benchmarks, while Fig.~\ref{fig:inverse32_branches} shows that they lie on continuous bound-state branches selected by the spectral Wronskian and that the polynomial degree $N$ does not label a unique spectral branch. Polynomiality therefore identifies special algebraic points of a bound-state problem whose physical selection is set by the global endpoint condition.

This distinction becomes physically sharper at the continuum edge. The inequality $U_0a>4$ follows from the first polynomial termination condition, whereas the exact threshold condition~\eqref{eq:critical_couplings} gives channel-dependent critical couplings below $4$. As illustrated in Fig.~\ref{fig:inverse32_thresholds}, the threshold sequences depend on both spin and angular momentum, so varying $U_0a$ changes the number of negative-energy states differently in different $(m,\tau)$ channels. {This channel dependence has a direct physical origin in the effective Hamiltonian~\eqref{eq:effective_hamiltonian}.}

{
For $U_0a<4$, the long-range contribution $+(4-U_0a)/\xi$ is repulsive; however, when $4m+\tau<0$, the combined orbital--Zeeman term $(4m+\tau)/\xi^{3/2}$ is attractive. This channel-dependent attraction can overcome the repulsive long-range contribution over the relevant radial region and support normalizable states in selected $(m,\tau)$ sectors, with their onset fixed by Eq.~\eqref{eq:critical_couplings}. The resulting spin--angular-momentum selectivity therefore reflects this competition between the radial contributions and is selected by the global bound-state condition rather than by polynomial truncation.
}

Downing and Portnoi emphasized that modulation of the magnetic-field strength can exclude vortexlike states with specific angular momenta and spin orientations, suggesting localization--delocalization transitions and a mechanism for spin separation~\cite{PRB94:045430:2016}. The present results place that physical picture on a global spectral footing for the inverse-$3/2$ model.

{
To place these results in a broader magnetic-confinement context, Table~\ref{tab:magnetic_comparison} compares the present model with representative magnetic confinement systems, emphasizing the magnetic-field profile, bound-state content, confinement criterion, spin- and angular-momentum dependence, and solution approach.
}

\begin{table*}[t]
\centering
\caption{Comparison of representative magnetic-confinement models and the present inverse-$3/2$ system.}
\label{tab:magnetic_comparison}
{\scriptsize
\renewcommand{\arraystretch}{1.18}
\setlength{\tabcolsep}{3.5pt}
\begin{tabularx}{\textwidth}{@{}p{0.125\textwidth} p{0.155\textwidth} X p{0.175\textwidth} p{0.17\textwidth} p{0.16\textwidth}@{}}
\toprule
Model / reference & Magnetic profile & Bound-state content & Critical confinement condition & Spin / angular-momentum selectivity & Solution approach \\
\midrule
Reijniers \textit{et al.}~\cite{PRB59:2817:1999} & Nonuniform field generated by a superconducting disk, producing a magnetic-antidot geometry & Magnetically confined bound states; dot filling changes discretely with magnetic-field strength & No universal analytic critical coupling reported; confinement depends on the detailed field profile and field strength & Angular-momentum-dependent states; no explicit spin-selective confinement analyzed & Numerical single-particle spectral calculation \\
\addlinespace
Ramezani Masir \textit{et al.}~\cite{PRB79:155451:2009} & Circular step magnetic-field profile & No true bound states; long-lived quasibound states occur near Landau levels & No bound-state threshold for this short-range magnetic profile & Quasibound spectrum resolved by angular momentum; no explicit spin-selective confinement & Radial Schr\"odinger/Dirac analysis and quasibound-state characterization \\
\addlinespace
Downing and Portnoi~\cite{PRB94:045430:2016} & Slowly decaying inverse-$3/2$ magnetic field with Coulomb attraction & Isolated exact/QES bound states; global bound-state spectrum not determined & Polynomial/QES sector subject to $U_0a>4$ & Explicit dependence on $m$ and spin $\tau$; channel-selective localization & Exact and quasi-exact Heun-polynomial construction \\
\addlinespace
Kuru \textit{et al.}~\cite{JPCM30:365502:2018} & Axially symmetric graphene magnetic fields with $B(r)\sim r^\alpha$ & Infinite zero modes and excited states for $\alpha\geq-1$; infinite zero modes only for $-2\leq\alpha<-1$; finite zero modes for $\alpha<-2$ & Confinement controlled by the asymptotic decay exponent, with boundaries at $\alpha=-1$ and $\alpha=-2$ & Dependence on $j$, combining angular momentum and isospin & Dirac--Weyl equation, supersymmetric quantum mechanics, and Aharonov--Casher analysis \\
\addlinespace
Present work & Inverse-$3/2$ magnetic field with Coulomb attraction & Global nonpolynomial bound-state branches; exact channel-resolved state count in the subcritical region & $(U_0a)_{c,n_r}^{(m,\tau)}=4-(4m+\tau)^2/(n_r+2|m|+1/2)^2$ & Explicit spin- and $m$-dependent thresholds and bound-state counts & Global BHE spectral Wronskian and exact continuum-edge analysis \\
\bottomrule
\end{tabularx}
}
\end{table*}

{
As summarized in Table~\ref{tab:magnetic_comparison}, magnetic confinement can take qualitatively different forms depending on the field profile and on the underlying carrier dynamics. The present inverse-$3/2$ model is distinguished by the availability of explicit spin- and angular-momentum-dependent confinement thresholds together with a global spectral condition beyond the QES sector.
}

A related global quantization strategy has earlier precedents in the BHE literature. Batola~\cite{ARMA78:275:1982,AM20:87:1982} derived connection relations and integral representations for the asymptotic coefficient of the solution regular at the origin, and Karwowski and Witek~\cite{TCA133:1494:2014} used the vanishing of the corresponding coefficient, $\mathcal K=0$, as the quantization condition for harmonium-type problems. Their implementation expresses the coefficient through integrals involving the BHE function; the absolute-convergence domain is described as rather narrow, and they identify tractable algorithms for square-integrable nonpolynomial states as an outstanding challenge~\cite{TCA133:1494:2014}. Here the same global spectral requirement is implemented differently: the condition is written directly as the vanishing of the Wronskian between the endpoint-selected Frobenius solution and the subdominant Thom\'e solution at infinity, and the required connection coefficient is evaluated through the construction of Ferreira and Sesma~\cite{NA71:797:2016}, based on the Heaviside-series method of Naundorf~\cite{SIAM7:157:1976}. This formulation avoids direct evaluation of the integral representation and provides a scalar spectral condition that can be followed continuously in parameter space. In the inverse-$3/2$ model, this makes it possible to retain the QES solutions as exact benchmarks while tracing generic nonpolynomial branches and their confinement thresholds.

{The present analysis concerns the $\mathcal E<0$ bound-state sector and the $\mathcal E=0$ continuum threshold of the two-dimensional Schr\"{o}dinger--Pauli model with the ideal inverse-$3/2$ magnetic profile, attractive Coulomb interaction, integer $m$, and $\tau=\pm1$.} Because the inverse-$3/2$ magnetic profile is singular at the origin, it is most naturally regarded as an analytically controlled model of confinement and delocalization rather than as a quantitative description of a regular experimental field profile. {The present treatment does not include a regularized finite-core magnetic profile or the positive-energy scattering sector.} A natural extension is therefore to regularize the field near the origin and determine how the bound-state branches, critical couplings, and spin--angular-momentum selectivity evolve for a finite-core magnetic profile. Such a comparison would distinguish features tied to the analytically controlled inverse-$3/2$ model from those that persist for a physically regular short-distance field. A second natural extension is to explore the positive-energy sector, where the same global BHE framework can be used to investigate scattering and its relation to the bound-state thresholds identified here.

\section{Conclusions}
\label{sec:conclusions}

In this work, we determined the global bound-state spectrum of a two-dimensional electron governed by the Schr\"{o}dinger--Pauli equation in an inverse-$3/2$ radial magnetic-field profile. {Treating the regularity and decay conditions as a global boundary-value problem shows that the quasi-exact polynomial solutions represent special members of a broader discrete spectrum rather than the full set of admissible states. Accordingly, the polynomial degree $N$ remains a termination index within the QES sector but does not provide a global labeling of the spectrum. The continuum-edge analysis also gives explicit channel-dependent critical couplings and bound-state counts, establishing that confinement may occur for $U_0a<4$ in selected spin--angular-momentum channels.}

{Beyond this specific magnetic model, the principal methodological result is a global BHE spectral criterion that is independent of polynomial termination. The same strategy can, in principle, be transferred to other eigenvalue problems reducible to a biconfluent Heun equation with well-defined endpoint boundary conditions. The present treatment is limited to the bound-state and continuum-threshold sectors of the ideal singular inverse-$3/2$ profile; finite-core regularization and extension to positive-energy scattering therefore provide natural tests of the robustness and broader scope of the results.}

\section*{Funding}
This work was supported in part by means of funds provided by CNPq, Brazil, Grant Nos. 308172\slash 2023-0 and CNPq\slash Universal\slash 420896/2025-2, FAPEMA, and CAPES (Finance Code 001).

\section*{Declaration of competing interest}
The authors declare that they have no known competing financial interests or personal relationships that could have appeared to influence the work reported in this paper.

\section*{Acknowledgements}
The author thanks the anonymous reviewers for their careful reading of the manuscript and for their constructive comments and suggestions, which helped improve its clarity and presentation.

\section*{Data availability}
No data was used for the research described in the article.




\bibliographystyle{elsarticle-num} 






\end{document}